\documentclass[aip,jcp,reprint]{revtex4-1}
\usepackage{graphicx}
\usepackage{color}
\usepackage[caption=false]{subfig}
\usepackage{amsmath,amssymb}
\usepackage{float}
\usepackage[utf8]{inputenc}
\usepackage{mathptmx}
\usepackage{soul}
\usepackage{mathtools}

\begin{document}
\title{Thermal feedback as a kinetic control mechanism in reaction–diffusion pattern formation}
\author{Sudip Dutta}
\affiliation{School of Chemistry, Indian Institute of Science Education and Research Thiruvananthapuram, Kerala, 695551, India.}

\author{Pushpita Ghosh}
\email{pushpita@iisertvm.ac.in}
\affiliation{School of Chemistry, Indian Institute of Science Education and Research Thiruvananthapuram, Kerala, 695551, India.}
\affiliation{Center for High-Performance Computing, Indian Institute of Science Education and Research, Thiruvananthapuram, Kerala 695551, India.}

\begin{abstract}
Pattern formation in reaction-diffusion systems is traditionally analyzed under isothermal assumptions, overlooking the dynamical role of temperature in systems where reactions generate and dissipate heat. Here, we investigate non-isothermal reaction-diffusion dynamics by coupling activator-inhibitor kinetics to a dynamically evolving temperature field that modulates reaction rates through Arrhenius-type dependencies. This coupling introduces an additional feedback mechanism that influences stability and pattern selection. Through analytical and numerical analysis of the Cholrine dioxide-Iodine-Malonic acid (CDIMA) and Schnakenberg models, we demonstrate that thermal feedback modifies dispersion relations by enhancing instability growth rates and shifting pattern selection toward shorter wavelengths. Beyond these intrinsic effects, we identify a boundary-mediated mechanism in which thermal constraints qualitatively alter global dynamics. In particular, fixed-temperature boundaries induce nonstationary behavior in the CDIMA system, whereas the Schnakenberg model exhibits robust stationary patterns. These results establish thermal-kinetic coupling as a general mechanism for controlling pattern formation and highlight the role of boundary-mediated heat exchange as a tunable parameter for spatiotemporal organization.

 \end{abstract}

\maketitle

\section{Introduction}
Spontaneous pattern formation in reaction--diffusion systems is a hallmark of nonequilibrium chemical and biological processes~\cite{Turing1952,Cross1993,epstein1998introduction}. 
Since Turing’s seminal proposal of diffusion-driven instability~\cite{Turing1952}, it has been understood that spatial symmetry breaking can arise purely from the interplay of local reactions and transport. Subsequent developments in nonequilibrium thermodynamics established such structures as dissipative states sustained far from equilibrium~\cite{Nicolis1977}.

\par Experimental realizations in chemical systems, most notably in the Belousov-Zhabotinsky and chlorite-iodide-malonic acid reactions, confirmed that reaction--diffusion mechanisms can generate stationary and oscillatory spatial structures~\cite{Zaikin1970,Castets1990,Epstein1996,Lengyel1991,Dulos1996}. These systems provide controllable platforms to investigate the kinetic and transport parameters governing instability thresholds and wavelength selection. Moreover, the emergence of spatial and spatiotemporal structures has been extensively studied in systems where local reaction kinetics are coupled to molecular diffusion, revealing a rich diversity of stationary and dynamical patterns governed by external perturbations such as light, noise, electric fields, temperature gradient and advective transport~\cite{horvath_1999,dolnik_2001,ghosh_2011_periodic,schmidt_2003,
nagao_2013,ghosh_2019,Tarpan2022,Tarpan2023,Tarpan2025}. Beyond chemistry, reaction-diffusion principles have become central to understanding biological morphogenesis and intracellular organization~\cite{Kondo2010}. Recent theoretical advances emphasize that pattern formation must be viewed as a dynamic, nonlinear mode-selection process shaped by conservation laws and feedback structure~\cite{Halatek2018}. In parallel, reaction-diffusion concepts increasingly inform materials self-assembly and programmable chemical systems~\cite{Grzybowski2009}.

\par Despite this progress, most theoretical and computational studies of reaction-diffusion pattern formation assume isothermal conditions, effectively treating temperature as a fixed and externally imposed parameter. This approximation is often justified for systems with efficient thermal equilibration. However, in many realistic chemical environments including exothermic reactions, confined reactors, and systems with limited heat dissipation, temperature evolves dynamically and is intrinsically coupled to the reaction network~\cite{Gray1984,Zeldovich1985}. In such systems, heat generation, diffusion, and exchange with the environment introduce an additional field that feeds back into reaction kinetics, potentially modifying instability thresholds, oscillatory behavior, and pattern selection.  

\par Previous studies of non-isothermal reaction-diffusion systems have demonstrated that temperature-dependent reaction rates can shift instability thresholds, modify oscillatory behavior, and influence pattern morphology~\cite{Serna,Silva-Dias,VanGorder2020}. However, a systematic understanding of how thermal feedback, in conjunction with boundary-mediated heat exchange, affects diffusion-driven instabilities and pattern selection remains incomplete. In particular, it remains unclear whether temperature acts merely as a passive rescaling of kinetic parameters or constitutes an independent control variable capable of qualitatively altering instability mechanisms.

\par In this work, we address these questions by developing and analyzing a thermally coupled reaction-diffusion framework in which temperature evolves as a dynamical field and modulates reaction kinetics through Arrhenius-type dependencies. The model consists of two interacting chemical species coupled to a heat equation that accounts for both internal heat generation and boundary-mediated thermal exchange. This formulation allows us to investigate the interplay between reaction kinetics, diffusion, and heat transport within a unified framework.

\par Our analysis reveals a key and nontrivial result: despite being a dynamical field, temperature does not introduce an independent unstable mode at the linear level. Instead, its effect can be understood as a state-dependent renormalization of reaction kinetics, which modifies instability growth rates and wavelength selection without altering the fundamental structure of the dispersion relation. In contrast, in the nonlinear regime, thermal boundary conditions play a decisive role by generating spatial temperature gradients that feed back into reaction rates, leading to qualitatively distinct dynamical behaviors. To demonstrate the generality of these effects, we analyze two prototypical activator-inhibitor systems: the CDIMA model and the Schnakenberg kinetics. We show that thermal coupling universally enhances instability growth and shifts pattern selection toward shorter wavelengths. However, the impact on global dynamics is strongly model dependent. In the CDIMA system, thermal boundary constraints can induce nonstationary behavior, whereas the Schnakenberg model retains robust stationary patterns under analogous conditions.

These results establish thermal-kinetic coupling as a general mechanism for controlling pattern formation in nonequilibrium systems and highlight the critical role of boundary-mediated heat exchange in determining spatiotemporal organization. More broadly, they demonstrate that temperature can act not only as a passive parameter but as an active control variable whose effects depend sensitively on the interplay between reaction nonlinearities, transport processes, and boundary conditions.

\section{Model and Methods}

To investigate the role of temperature as a critical control parameter in the emergence and stability of self-organized patterns, we consider a thermally coupled reaction–diffusion system in two spatial dimensions. Specifically, we focus on a generic autocatalytic chemical model involving intermediate species, in which the reaction-diffusion dynamics are coupled to an energy balance equation governing the evolution of the temperature field through temperature-dependent reaction rates \cite{VanGorder2020}. 

The governing equations are given by
\begin{equation}
\dot{u}= D_u\nabla^2 u + h_u(T) f(u,v), \tag{1a}\label{Equation:1a}
\end{equation}
\begin{equation}
\dot{v}= D_v\nabla^2 v + h_v(T) g(u,v), \tag{1b}\label{Equation:1b}
\end{equation}
\begin{equation}
\dot{T}= D_T \nabla^2 T - h_u(T) f(u,v) - h_v(T) g(u,v). \tag{1c}\label{Equation:1c}
\end{equation}

Here, $u$ and $v$ denote dimensionless concentration fields of two  intermediate species, while $T$ represents the temperature field. The nonlinear functions $f(u,v)$ and $g(u,v)$ describe the local reaction kinetics and are chosen to follow an activator-inhibitor mechanism. The functions $h_u(T)$ and $h_v(T)$ represent temperature-dependent scaling factors that modulate the reaction rates associated with species $u$ and $v$, respectively. The parameters $D_u$ and $D_v$ are the diffusion coefficients of the chemical species, and $D_T$ is the thermal diffusivity. The spatial domain is denoted by $x \in \Omega$, where $\Omega$ is a connected two-dimensional region, and time is defined for $t \geq 0$.

The temperature dependence of the reaction rates is modeled using an Arrhenius-type relation, which is commonly employed in non-isothermal reaction–diffusion systems \cite{molecules2021,Laidler1984TheDO}
\begin{equation}
h(T) = A \exp\left(-\frac{B}{T}\right), \tag{2}\label{Equation:2}
\end{equation}
where $A$ is the pre-exponential factor controlling the overall reaction strength, and $B$ is a parameter proportional to the activation energy, determining the sensitivity of the reaction rate to temperature variations. Accordingly, the temperature-dependent rate prefactors for the two species are given by
\begin{equation}
h_u(T) = A_u \exp\left(-\frac{B_u}{T}\right), \tag{3a}\label{Equation:3a}
\end{equation}
\begin{equation}
h_v(T) = A_v \exp\left(-\frac{B_v}{T}\right). \tag{3b}\label{Equation:3b}
\end{equation}
\begin{figure}[t]
    \centering
    \includegraphics[width=0.5\textwidth]{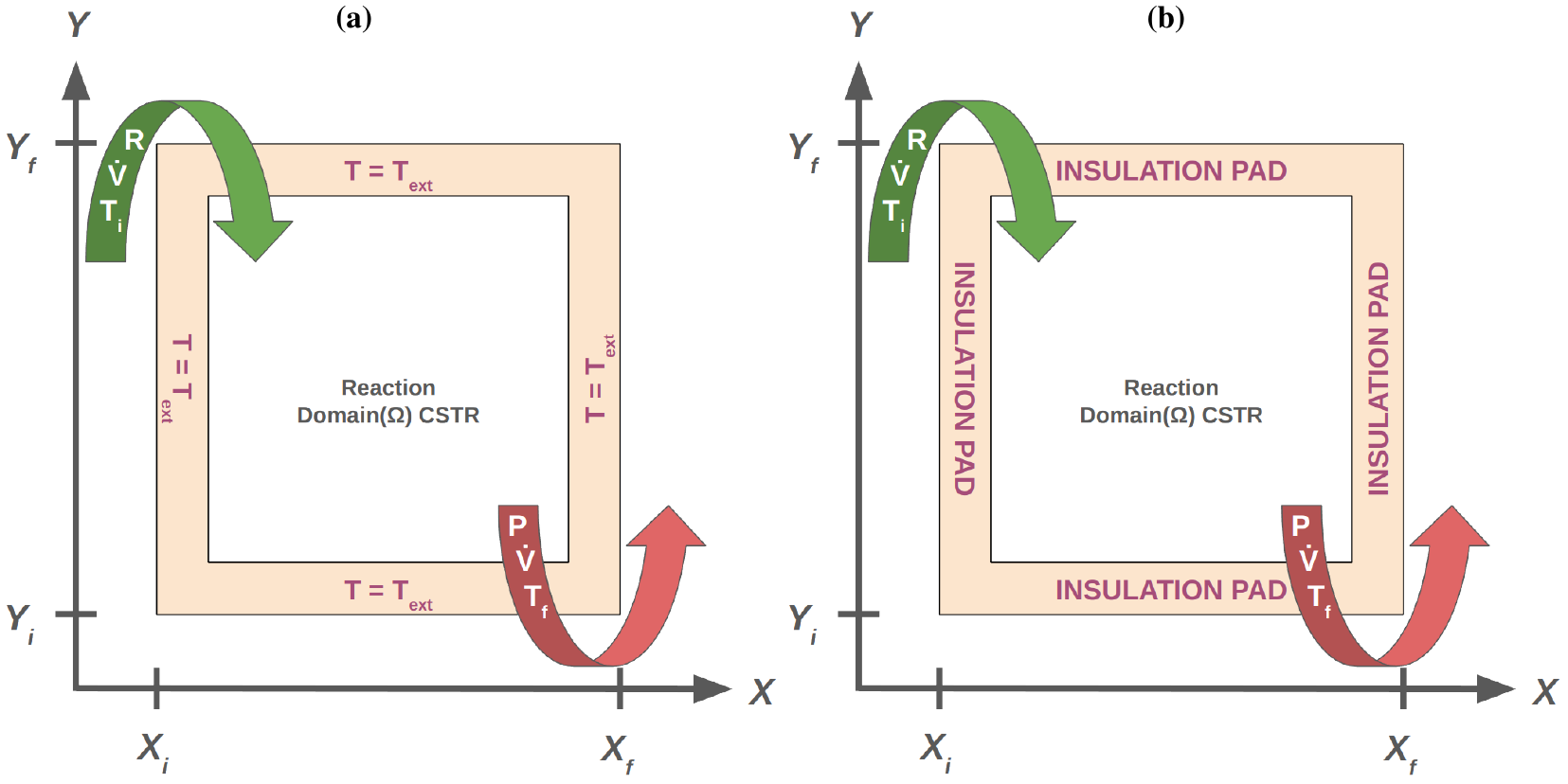}
    \caption{
Schematic illustration of the two-dimensional reaction domain showing $(a)$ Dirichlet and $(b)$ Neumann boundary conditions. $R$ and $P$ denote the reactant and product, respectively; $\dot{V}$ represents the velocity field; $T_i$ and $T_f$ denote the initial and final temperatures, respectively; and $T_{\mathrm{ext}}$ represents the external temperature.
}
    \label{fig:Figure1}
\end{figure}

\subsection*{Boundary Conditions}

The spatial domain $\Omega$ may possess a boundary $\partial\Omega$. For the chemical species $u$ and $v$, no-flux (Neumann) boundary conditions are imposed,
\[
\mathbf{n} \cdot \nabla u = \mathbf{n} \cdot \nabla v = 0\tag{4}\label{Equation:4} \quad \text{on } \partial\Omega,
\]
where $\mathbf{n}$ denotes the outward unit normal to the boundary. For the temperature field, heat exchange with the surroundings is modeled using Newton’s law of cooling,
\[
D_T\, \mathbf{n} \cdot \nabla T + \kappa (T - T_{\text{ext}}) = 0\tag{5}\label{Equation:5},
\]
where $\kappa$ is the heat transfer coefficient and $T_{\mathrm{ext}}$ denotes the external (bath) temperature. Two limiting cases of this boundary condition are of particular interest:
\begin{itemize}
\item \textbf{Condition 1 (Thermally insulated boundary):}
When $\kappa \ll D_T$, heat exchange with the environment is negligible, and the boundary condition effectively reduces to a Neumann condition, $\mathbf{n} \cdot \nabla T = 0$.
\item \textbf{Condition 2 (Isothermal boundary):}  
When $\kappa \gg D_T$, the boundary temperature is clamped to the external bath, yielding a Dirichlet condition $T = T_{\mathrm{ext}}$.
\end{itemize}

A schematic representation of the two-dimensional reaction domain and the corresponding thermal boundary conditions is shown in Fig.~\ref{fig:Figure1}. The figure illustrates the coupling between reaction-diffusion dynamics in the bulk and heat exchange at the system boundary, highlighting the two limiting cases of thermally insulated (Neumann) and isothermal (Dirichlet) conditions considered in this work. This setup serves as a minimal experimental or reactor-level realization of a non-isothermal reaction-diffusion system.

\subsection*{Linear Stability Analysis}
To determine how thermal coupling influences the onset of pattern formation, we analyze the linear stability of the spatially homogeneous steady state ($(u_{ss}, v_{ss}, T_{ss})$ ) with respect to small perturbations. While the steady-state concentrations ($u_{ss}$) and ($v_{ss}$) are determined by the reaction kinetics, the steady-state temperature ($T_{ss}$) is selected by the imposed thermal boundary conditions.
  
We first examine the temporal stability of the homogeneous steady state
$(u_{ss},v_{ss},T_{ss})$. Small perturbations around the steady state are
introduced as
\begin{eqnarray}
u=u_{ss}+\tilde{u}e^{\lambda t}, \quad
v=v_{ss}+\tilde{v}e^{\lambda t}, \quad
T=T_{ss}+\tilde{T}e^{\lambda t}.
\end{eqnarray}

The Jacobian matrix associated with the reaction kinetics of the variables
$u$ and $v$ is

\begin{equation}
J=
\begin{pmatrix}
\dfrac{\partial f}{\partial u} & \dfrac{\partial f}{\partial v} \\
\dfrac{\partial g}{\partial u} & \dfrac{\partial g}{\partial v}
\end{pmatrix}
=
\begin{pmatrix}
J_{11} & J_{12} \\
J_{21} & J_{22}
\end{pmatrix}.
\end{equation}

In the presence of temperature-dependent rate prefactors ($h_u(T)$) and ($h_v(T)$), the full linearized system takes the form
\begin{equation}
J_T=
\begin{pmatrix}
h_u(T_{ss})J_{11} & h_u(T_{ss})J_{12} & 0 \\
h_v(T_{ss})J_{21} & h_v(T_{ss})J_{22} & 0 \\
-h_u(T_{ss})J_{11}-h_v(T_{ss})J_{21} &
-h_u(T_{ss})J_{12}-h_v(T_{ss})J_{22} & 0
\end{pmatrix}.
\end{equation}
This matrix has a block-triangular structure, implying that one eigenvalue is identically zero, corresponding to a neutral homogeneous temperature mode. Consequently, the stability of the reactive subsystem is governed by the reduced Jacobian
\begin{equation}
\widehat{J}_T=
\begin{pmatrix}
h_u(T_{ss})J_{11} & h_u(T_{ss})J_{12} \\
h_v(T_{ss})J_{21} & h_v(T_{ss})J_{22}
\end{pmatrix}.
\end{equation}

The eigenvalues ($\lambda$) satisfy

\begin{eqnarray}
\lambda^2 - \lambda  \mathrm{tr}(\hat{J}_T) + \det(\hat{J}_T) = 0
\end{eqnarray}

%
%
%


The homogeneous steady state is temporally stable when

\begin{equation}
\mathrm{tr}(\widehat{J}_T)
=
h_u(T_{ss})J_{11}
+
h_v(T_{ss})J_{22}
<0,
\end{equation}

\begin{equation}
\det(\widehat{J}_T)
=
h_u(T_{ss})h_v(T_{ss})\det(J)
>0.
\end{equation}
To analyze spatiotemporal instabilities, we consider perturbations of the form
\begin{eqnarray}
u &=& u_{ss}+\tilde{u}e^{\lambda t+i\mathbf{q}\cdot\mathbf{r}},\qquad\\
v &=& v_{ss}+\tilde{v}e^{\lambda t+i\mathbf{q}\cdot\mathbf{r}},\qquad\\
T &=& T_{ss}+\tilde{T}e^{\lambda t+i\mathbf{q}\cdot\mathbf{r}}.
\end{eqnarray}

Linearizing the governing equations around the steady state yields the
Jacobian matrix in the presence of diffusion
{\begin{equation}
J_D=
\begin{pmatrix}
h_u(T_{ss})J_{11}-D_uq^2 & h_u(T_{ss})J_{12} & 0 \\
h_v(T_{ss})J_{21} & h_v(T_{ss})J_{22}-D_vq^2 & 0 \\
-h_u(T_{ss})J_{11}-h_v(T_{ss})J_{21} &
-h_u(T_{ss})J_{12}-h_v(T_{ss})J_{22} & -D_Tq^2
\end{pmatrix}.
\end{equation}}
The growth rates are determined from

\begin{equation}
\det(J_D-\lambda I)=0.
\end{equation}
To clarify how temperature enters the stability problem, we examine its effect on the effective reaction kinetics before proceeding to the full dispersion relation.

\subsubsection{Effective kinetic renormalization induced by temperature}

Although the governing equations explicitly couple the concentration fields to a dynamically evolving temperature field, the influence of temperature on linear stability can be understood in terms of \textit{effective kinetic renormalization}.

The reaction terms enter the model through temperature-dependent prefactors of Arrhenius form, $h_u(T)$ and  $h_v(T)$,
which multiplicatively scale the local reaction kinetics. Linearization about the homogeneous steady state $(u_{ss}, v_{ss}, T_{ss})$ yields
\begin{equation}
\delta\big(h_u(T) f(u,v)\big)
=
h_u(T_{ss})\left(J_{11} \tilde u + J_{12} \tilde v\right)
+
h_u'(T_{ss}) f(u_{ss},v_{ss}) \tilde T,
\end{equation}
and similarly for the $v$-equation. Because the steady state satisfies
\begin{equation}
f(u_{ss}, v_{ss}) = 0, \quad g(u_{ss}, v_{ss}) = 0,
\end{equation}
the terms proportional to the temperature perturbation $\tilde T$ vanish identically. As a result, the temperature field does not directly feed back into the reactive subsystem at linear order.  Consequently, the Jacobian of the linearized system acquires a block-triangular structure, in which the temperature perturbation mode decouples from the concentration dynamics.

Since the temperature equation decouples linearly, the determinant
factorizes as
\begin{multline}
(\lambda+D_Tq^2)
\Big[
\lambda^2
-
(\mathrm{tr}\,\widehat{J}_D)\lambda
+
\det(\widehat{J}_D)
\Big]
=0
\end{multline}
with one  eigenvalue, $\lambda_T=-D_Tq^2<0$, which is always stable. 
This factorization shows that thermal diffusion does not explicitly influence the onset of linear instability. Instead, temperature enters the stability problem solely through effective kinetic renormalization of the reaction rates.
The remaining eigenvalues that determine the onset of instability are governed by a reduced $2 \times 2$ Jacobian  involving only the concentration fields.
   \begin{equation}
 \widehat{J}_D=
  \begin{pmatrix}
h_u(T_{ss})J_{11}-D_uq^2 & h_u(T_{ss})J_{12} \\
  h_v(T_{ss})J_{21} & h_v(T_{ss})J_{22}-D_vq^2
 \end{pmatrix}.
\end{equation}

The remaining two eigenvalues satisfy the following dispersion relation:

\begin{equation}
\lambda^2
-
(\mathrm{tr}\,\widehat{J}_D)\lambda
+
\det(\widehat{J}_D)
=0,
\end{equation}
with two eigen values
\begin{equation}
\lambda_{\pm}=\tfrac{1}{2}\big[\mathrm{tr}\widehat{J}_D)\pm
\sqrt{(\mathrm{tr}\widehat{J}_D))^2-4\det(\widehat{J}_D))}\big].
\end{equation}

Within this reduced description, thermal effects enter the linear stability problem exclusively through the \textit{effective kinetic prefactors}
\begin{equation}
\alpha_u = h_u(T_{ss}), \quad \alpha_v = h_v(T_{ss}),
\end{equation}
which renormalize the entries of the reaction Jacobian. 
This formulation shows that, at the level of linear stability, temperature acts not as an independent dynamical mode but as a parameter that modifies the relative strength of the reaction pathways.

\subsubsection{Effective kinetic asymmetry and thermal control}

Within the present framework, thermal effects enter the linear stability problem through the effective kinetic prefactors
\begin{equation}
\alpha_u = A_u \exp\!\left(-\frac{B_u}{T_{ss}}\right), \quad
\alpha_v = A_v \exp\!\left(-\frac{B_v}{T_{ss}}\right).
\end{equation}
These quantities renormalize the activator and inhibitor reaction channels at the homogeneous steady state. In the present analysis, the activation parameters $B_u$ and $B_v$ are kept fixed, while the pre-exponential factors $A_u$ and $A_v$ are used as the primary control parameters. Accordingly, the thermal modulation explored here is most directly interpreted as a change in the effective kinetic weighting of the two reaction pathways.

A useful measure of this relative weighting is the ratio
\begin{equation}
\Gamma = \frac{\alpha_u}{\alpha_v}
= \frac{A_u}{A_v}\exp\!\left(-\frac{B_u-B_v}{T_{ss}}\right).
\end{equation}
For fixed $B_u$ and $B_v$, variations in $A_u$, $A_v$, or $T_{ss}$ modify $\Gamma$ and thereby shift the effective balance between activator and inhibitor dynamics. Thus, within the linear regime, temperature acts as a control parameter through effective kinetic renormalization rather than through thermal diffusion directly.

Although the activation parameters $B_u$ and $B_v$ are not varied in the present study, their difference plays an important conceptual role. When $B_u \neq B_v$, the Arrhenius factors introduce an intrinsic asymmetry in the temperature sensitivity of the activator and inhibitor pathways. As a result, changes in temperature do not simply rescale the overall reaction rates but modify the relative strength of the two reaction channels.

This effect is reflected in the ratio $\Gamma$
which depends explicitly on temperature when $B_u \neq B_v$. Thus, temperature modifies the relative strength of activator and inhibitor pathways through differential Arrhenius scaling, providing a direct mechanism for tuning instability thresholds.

\subsubsection{Role of thermal diffusion and boundary conditions}

Although thermal diffusion does not explicitly enter the reduced dispersion relation, it plays an important role in determining the spatial structure of the temperature field and its interaction with the boundaries. In particular, the boundary heat transfer coefficient $\kappa$ governs whether temperature is locally determined by internal dynamics (Neumann limit) or constrained by the external environment (Dirichlet limit).

In the nonlinear regime, these boundary-induced temperature variations can generate spatially heterogeneous reaction rates, thereby influencing pattern selection and global dynamics even though thermal diffusion does not directly control the linear instability condition.

\vspace{0.5em}
\noindent
This separation between linear kinetic renormalization and nonlinear boundary-mediated effects is central to understanding the model-dependent behavior observed in the following sections.

\subsection*{Condition for spatial instability}
Diffusion-driven (Turing) instability occurs when a homogeneous steady state that is stable in the absence of diffusion becomes unstable for finite wavenumber ($q$). This requires
{\begin{multline}
\det(\widehat{J}_D)
=
q^4D_uD_v
-
q^2
\Big(
h_u(T_{ss})J_{11}D_v
+
h_v(T_{ss})J_{22}D_u
\Big)
+\\
h_u(T_{ss})h_v(T_{ss})\det(J)<0.\label{Eq.22}
\end{multline}}
The instability band is determined by the roots
\begin{equation}
q_{\pm}^{2}=\frac{1}{2D_uD_v}
\left[A \pm \sqrt{A^{2}-4D_uD_v\,h_u(T_{ss})h_v(T_{ss})\det(J)}\right],\label{Eq.23}
\end{equation}
where
\begin{equation}
A = h_u(T_{ss}) J_{11}D_v + h_v(T_{ss}) J_{22}D_u.
\end{equation}
A necessary condition for instability is
\begin{equation}
h_u(T_{ss})J_{11}D_v+h_v(T_{ss})J_{22}D_u>0.\label{Eq.24}
\end{equation}
together with the requirement that the discriminant~(in Eq.~\ref{Eq.23}) be positive. These conditions define the onset of diffusion-driven pattern formation.

\begin{figure}[t]
    \centering
    \includegraphics[width=0.5\textwidth]{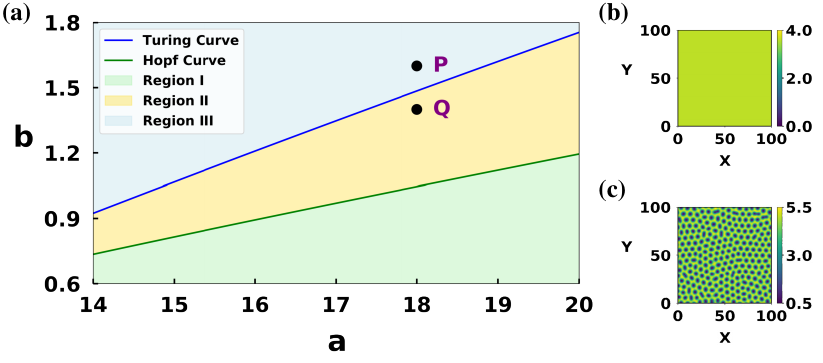}
    \caption{
$(a)$ Bifurcation diagram in the $a$--$b$ parameter space. Light green denotes the spatiotemporally unstable region, light yellow indicates spatial instability with temporal stability, and light blue represents the spatiotemporally stable region. Points $P(a=18,\; b=1.6)$ and $Q(a=18,\; b=1.4)$ denote representative parameter sets. 
$(b,c)$ Numerical simulation results corresponding to the representative points $P$ and $Q$, respectively. The parameters $\sigma = 9$ and $D = 1.6$ are fixed.
}\label{Fig:CDIMA_bif}
\end{figure}

\subsection*{Numerical methods}
The governing reaction–diffusion equations are solved using an explicit Euler scheme for time integration combined with a finite-difference discretization of spatial derivatives in two dimensions. The computations are performed on square domains with system sizes and discretization parameters chosen appropriately for each model (described in the next section) to ensure numerical stability and sufficient resolution of emerging spatial structures. 

Simulations are initiated from the homogeneous steady state with small-amplitude random perturbations (typically $\pm 1\%$) to break spatial symmetry and seed pattern formation. The time step is chosen to ensure numerical stability and convergence, and the spatial grid resolution is verified to resolve the shortest unstable wavelengths predicted by linear stability analysis.

\section{Results and Discussion}
We now investigate how coupling reaction-diffusion dynamics to a self-consistently evolving temperature field reshapes both homogeneous and diffusion-driven instabilities. The analysis distinguishes between two complementary effects of thermal coupling: (i) modification of instability thresholds through temperature-dependent kinetic renormalization, and (ii) boundary-mediated feedback that influences nonlinear pattern selection.

To demonstrate the generality of the proposed framework, we analyze two chemically distinct but canonical activator–inhibitor systems: the Chlorine dioxide–Iodine–Malonic acid (CDIMA) model and the activator-depleted Schnakenberg kinetics. For each system, we first establish the baseline (athermal) stability structure, which serves as a reference for interpreting the effects of thermal coupling. We then examine how temperature-dependent reaction rates, entering through Arrhenius-type prefactors, modify Hopf and Turing instabilities through temperature-induced renormalization of the underlying reaction kinetics. 
This is followed by an analysis of the dispersion relation and wavelength selection, highlighting how temperature controls both instability growth rates and intrinsic length scales. Finally, we investigate nonlinear pattern formation under different thermal boundary conditions, demonstrating how boundary-imposed thermal constraints can significantly alter global dynamics. Detailed derivations and extended parameter scans are provided in the Supporting Information (SI), while the main text focuses on the physical mechanisms through which thermal coupling modifies instability and pattern selection.

\subsection{MODEL -- I: CDIMA  System}
\subsubsection*{Baseline isothermal instability structure}
We begin by establishing the baseline isothermal stability properties of the CDIMA system, which serve as a reference for assessing the impact of thermal coupling. The dimensionless two-variable model introduced by Lengyel and Epstein~\cite{epstein1998introduction} reads:
\begin{align}
\frac{\partial u}{\partial t} &= f(u,v)_{\mathrm{CDIMA}} = a - u - \frac{4uv}{1 + u^2} + \nabla^2 u, \\
\frac{\partial v}{\partial t} &= g(u,v)_{\mathrm{CDIMA}} = \sigma \left[ b\left(u - \frac{uv}{1+u^2}\right) + D \nabla^2 v \right],
\end{align}
where $u$ and $v$ denote the dimensionless concentrations of the activator (I$^-$) and inhibitor (ClO$_2^-$), respectively. The parameters $a$ and $b$ are proportional to the initial reactant concentrations, $D$ is the ratio of diffusion coefficients, and $\sigma$ reflects the concentration of the complexing agent. 

The homogeneous steady state is given by $u_{\mathrm{ss}} = a/5$ and $v_{\mathrm{ss}} = 1 + a^2/25$. Linear stability analysis (see SI) yields the conditions for Hopf and diffusion-driven (Turing) instabilities. The resulting bifurcation diagram in the $(a,b)$ parameter space is shown in Fig.~\ref{Fig:CDIMA_bif}(a), delineating regions corresponding to homogeneous steady states, Turing instability, and Hopf instability. To illustrate these regimes, we select representative parameter sets denoted by points $P$ and $Q$. Point $P$ lies in the linearly stable homogeneous region and yields a spatially uniform steady state, whereas point $Q$ lies in the Turing-unstable region and produces a stationary spatial pattern, as shown in Figs.~\ref{Fig:CDIMA_bif}(b) and (c) respectively.
\begin{figure*}[t]
    \centering
    \includegraphics[width=0.95\textwidth]{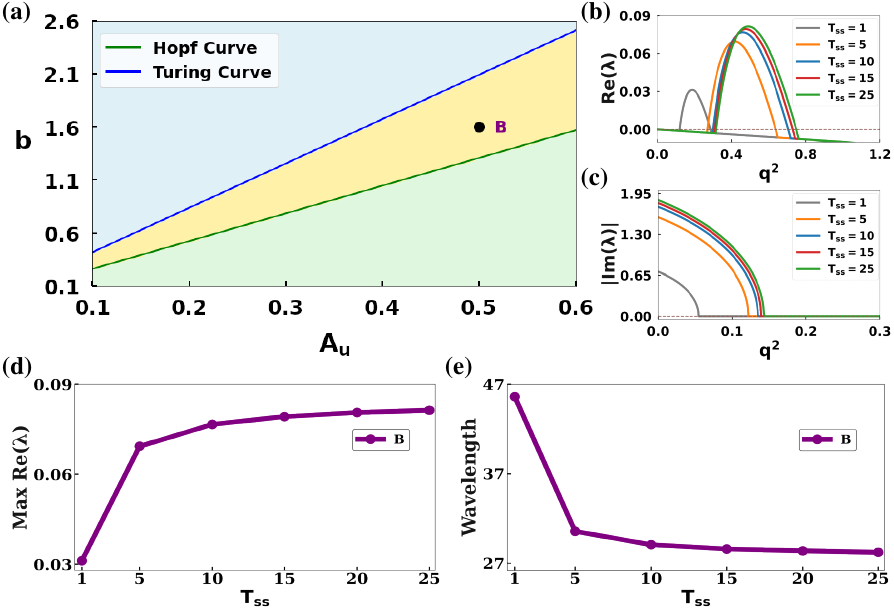}
     \caption{Temperature-modulated stability and pattern selection in the CDIMA reaction--diffusion system.
(a) Bifurcation diagram in the $A_u$--$b$ parameter space at fixed $a = 18$, $A_v = 0.4$, $\sigma = 9$, and $B_u = B_v = 1$. The green and blue curves denote the Hopf and Turing instability boundaries, respectively. The light green region corresponds to spatiotemporal instability, the light yellow region to purely spatial (Turing) instability, and the light blue region to the stable regime. Point $B$ ($0.5,1.6$) lies within the Turing region and is selected for further analysis.
(b,c) Dispersion relations evaluated at point $B$ for different steady-state temperatures $T_{ss}$: (b) real part $\mathrm{Re}(\lambda)$ and (c) imaginary part $\mathrm{Im}(\lambda)$ as functions of the wavenumber $q^2$. The positive $\mathrm{Re}(\lambda)$ over a finite band of $q^2$ and vanishing $\mathrm{Im}(\lambda)$ confirm a stationary Turing instability.
(d,e) Temperature dependence at point $B$: (d) maximum growth rate $\max[\mathrm{Re}(\lambda)]$ and (e) corresponding pattern wavelength. Increasing $T_{ss}$ enhances the growth rate while reducing the characteristic wavelength, indicating temperature-driven control of pattern amplitude and spatial scale.}
     \label{Fig:CDIMA_temp_bif}
\end{figure*}

Further insight into the instability mechanisms is obtained from the dispersion relations evaluated at representative points. In the Turing regime, the real part of the eigenvalue exhibits a positive peak at a finite wavenumber, indicating the growth of stationary spatial modes, while the imaginary part remains zero. In the Turing regime, the real part of the growth rate $\mathrm{Re}(\lambda)$ exhibits a positive maximum at a finite wavenumber, indicating the amplification of stationary spatial modes, while the imaginary part vanishes. In contrast, the Hopf regime is characterized by a nonzero imaginary component at zero wavenumber, corresponding to temporal oscillations without spatial structure.

\par Having established the baseline isothermal instability structure of the CDIMA system, we now examine how thermal coupling reshapes this landscape. As discussed in Sec.~II, temperature enters through Arrhenius-type prefactors that renormalize the effective reaction kinetics. Consequently, temperature acts as a control parameter that modifies the balance between activator and inhibitor dynamics and shifts the location of instability boundaries.
\begin{figure*}[t]
    \centering
    \includegraphics[width=0.95\textwidth]{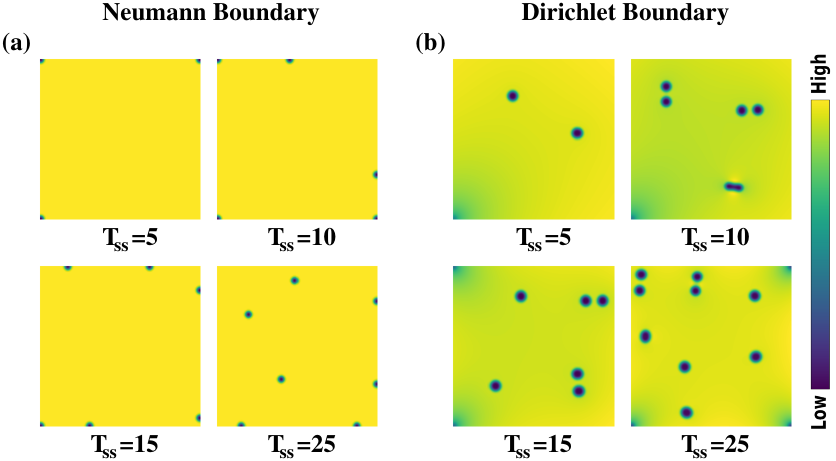}
  \caption{Numerically simulated spatial patterns corresponding to point $B$ in the $b$--$A_u$ bifurcation diagram.
(a) Pattern obtained under Neumann boundary conditions.
(b) Pattern obtained under Dirichlet boundary conditions.
Simulations are performed with diffusion coefficients $D_T = 0.01$ and $D = 1.6$ on a square domain of size $100 \times 100$ with spatial discretization $\Delta x = 0.5$ and time-step $dt=0.002$.}
  \label{Fig:CDIMA_numerics}
\end{figure*}
\subsubsection*{Thermal renormalization of instability in the CDIMA system}
To incorporate thermal effects, we introduce Arrhenius-type temperature dependence into the reaction terms. The temperature-coupled reaction–diffusion system is given by
\begin{align}
\frac{\partial u}{\partial t} &= \nabla^2 u + h_u(T)f(u,v)_{\mathrm{CDIMA}}, \\
\frac{\partial v}{\partial t} &= \sigma D \nabla^2 v + h_v(T)g(u,v)_{\mathrm{CDIMA}}, \\
\frac{\partial T}{\partial t} &= D_T \nabla^2 T - h_u(T)f(u,v)_{\mathrm{CDIMA}} - h_v(T)g(u,v)_{\mathrm{CDIMA}},
\end{align}
where $h_u(T) = A_u \exp(-B_u/T)$ and $h_v(T) = A_v \exp(-B_v/T)$ denote the temperature-dependent kinetic prefactors associated with the activator and inhibitor pathways, respectively. 

\par Linear stability analysis about the homogeneous steady state yields the bifurcation structure in the $(A_u, b)$ parameter space as illustrated in Fig.~\ref{Fig:CDIMA_temp_bif}(a) (see SI for details). The Hopf bifurcation boundary is given by
\begin{eqnarray}
A_u < -A_v \, e^{\frac{B_u - B_v}{T_{ss}}} \,
\left[\frac{-5ab\sigma}{(3a^2-125)}\right],
\end{eqnarray}
which separates temporally stable and oscillatory regimes. The corresponding Turing instability condition is obtained from the diffusion-driven instability criteria,
\begin{widetext}
\begin{equation}
\left[
D A_u e^{-\frac{B_u}{T_{ss}}}
\frac{3u_{ss}^2 - 5}{1 + u_{ss}^2}
+  A_v e^{-\frac{B_v}{T_{ss}}}
\frac{\sigma(-b)u_{ss}}{1 + u_{ss}^2}
+ 2 \sqrt{
D
\left(
\frac{3u_{ss}^2 - 5}{1 + u_{ss}^2}
\cdot
\frac{\sigma(-b)u_{ss}}{1 + u_{ss}^2}
-
\frac{-4u_{ss}}{1 + u_{ss}^2}
\cdot
\frac{2\sigma b u_{ss}^2}{1 + u_{ss}^2}
\right)
A_u A_v e^{-[\frac{B_u+B_v}{T_{ss}}]}
}
\right] > 0
\end{equation}
\end{widetext}
which defines the boundary for the onset of stationary spatial patterns.

These results show that thermal coupling shifts the instability boundaries through temperature-dependent renormalization of the effective kinetic parameters. In particular, variations in $A_u$ modify the effective activator strength $\alpha_u = A_u e^{-B_u/T_{ss}}$, thereby altering the relative balance between activator and inhibitor dynamics and displacing the system within the underlying kinetic phase space.

\par When $B_u = B_v$, the bifurcation conditions become independent of temperature, and thermal effects enter only through the effective prefactors. To isolate this effect, we analyze the dispersion relation at a representative point $B$ in the bifurcation diagram. The real and imaginary parts of the eigenvalues are shown as functions of $q^2$ in Fig.~\ref{Fig:CDIMA_temp_bif}(b,c) for different temperatures.
As temperature increases, the growth rate of unstable modes, $\mathrm{Re}(\lambda)$, increases monotonically, indicating enhanced instability. Concurrently, the most unstable mode shifts toward higher wavenumbers, implying a reduction in the characteristic wavelength of the emerging patterns. This behavior reflects the temperature-induced amplification of reaction kinetics, which strengthens local reactive processes relative to diffusion.

To quantify this trend, we examine the maximum growth rate $\max{\mathrm{Re}(\lambda)}$ and the corresponding dominant wavelength as functions of temperature [Fig.~\ref{Fig:CDIMA_temp_bif}(d)]. The growth rate increases with temperature, with a tendency toward saturation at higher values, while the wavelength decreases monotonically. This demonstrates a systematic transition from coarse to finer spatial structures. Taken together, these results show that temperature not only shifts instability thresholds but also provides direct control over both the growth rate and intrinsic length scale of pattern formation.

When the activation barriers differ ($B_u \neq B_v$), the ratio $\Gamma$ introduces an additional temperature-dependent pathway, thereby coupling the dynamics more strongly to the thermal field. As a consequence, the steady-state temperature $T_{\mathrm{ss}}$ explicitly enters the bifurcation structure and modifies the instability boundaries. We construct a three-parameter bifurcation diagram in terms of $b$, $A_u$, and $T_{\mathrm{ss}}$ (Fig.~S1 in SI), which delineates the regions of Hopf and Turing instabilities. We then analyze four representative points ($B_1-B_4$) within the Turing regime, where the real part of the leading eigenvalue is positive. In all cases, the most unstable mode shifts toward higher wavenumbers, indicating a reduction in the characteristic length scale of the emerging patterns.
\begin{figure}[t]
\centering
\includegraphics[width=0.5\textwidth]{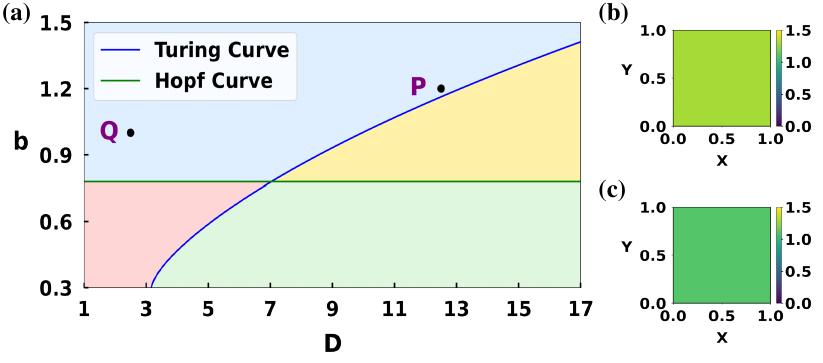}
\caption{
$(a)$ Bifurcation diagram in the $b$--$D$ parameter space. Light green denotes the spatiotemporally unstable region; light yellow indicates spatial instability with temporal stability; light blue represents the spatiotemporally stable region; and light pink corresponds to spatially stable but temporally unstable behavior. Points $P(D=12.5,\; b=1.2)$ and $Q(D=2.5,\; b=1.0)$ denote representative parameter sets. 
$(b,c)$ Numerical simulation results corresponding to the representative points $P$ and $Q$, respectively. The parameter $a=0.1$ is fixed.
}
\label{Fig:Schnak_bif}
\end{figure}

\begin{figure*}[t]
\centering
\includegraphics[width=0.95\textwidth]{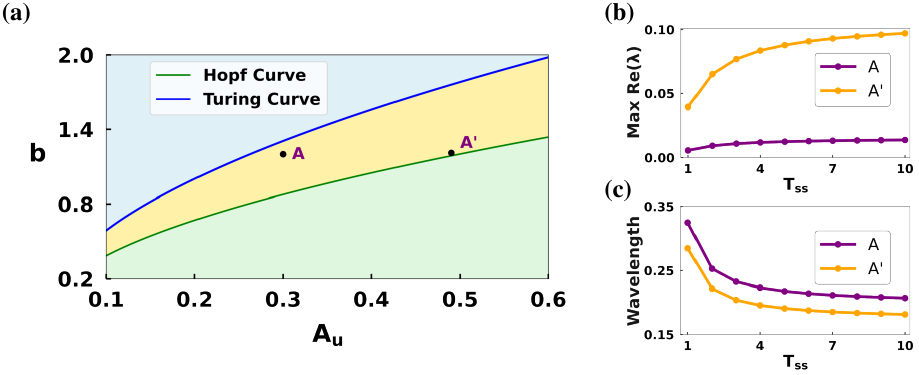}
\caption{
$(a)$ Two-parameter bifurcation diagram of the temperature-coupled Schnakenberg model in the $A_u$--$b$ parameter space for a representative homogeneous steady-state point $P(D=12.5,\; b=1.2)$. Light green denotes the spatiotemporally unstable region, light yellow the spatially unstable but temporally stable region, and light blue the spatiotemporally stable region. 
$(b,c)$ Maximum growth rate $\max \mathrm{Re}(\lambda)$ and the corresponding pattern wavelength as functions of temperature for the points $A(0.30,1.20)$ and $A^{\prime}(0.49,1.20)$, respectively.
}
\label{Fig:Schnak_bif_P}
\end{figure*}
\subsubsection*{Spatial patterns and boundary-condition effects}

To complement the analytical predictions through linear stability analysis, we numerically simulate the spatially extended temperature-coupled CDIMA system under zero-flux (Neumann) boundary conditions for all fields. These conditions ensure that no mass or heat flux crosses the system boundaries, thereby isolating the intrinsic bulk pattern-forming dynamics. The resulting steady-state patterns for different initial temperatures are shown in Fig.~\ref{Fig:CDIMA_numerics}(a).


Consistent with the stability analysis, increasing temperature leads to the emergence of a larger number of spot-like inhomogeneities within the fixed spatial domain. This observation directly reflects the shift of the most unstable mode toward higher wavenumbers with temperature, as inferred from the dispersion relation. Accordingly, the characteristic wavelength decreases at elevated temperatures, resulting in finer spatial structures.

To further assess the influence of boundary constraints, we perform simulations in which Dirichlet boundary conditions are imposed only on the temperature field $T$, while the concentration fields $u$ and $v$ continue to satisfy Neumann (zero-flux) conditions. The comparison is presented in Fig.~\ref{Fig:CDIMA_numerics}(b).
Under this mixed boundary setup, the spots appear larger, become more closely spaced, and exhibit reduced spatial periodicity across the domain. Since the chemical species remain flux-free, this qualitative change arises solely from fixing the temperature at the boundaries. The imposed constant temperature effectively acts as a thermal reservoir, generating spatial thermal gradients that influence local reaction rates through the Arrhenius factors $h_u(T)$ and $h_v(T)$. Thus, even when the chemical concentrations are not directly constrained at the boundaries, thermal Dirichlet conditions can significantly reshape the selected spatial modes.

Furthermore, a qualitative change in dynamical behavior is observed. Under Neumann boundary conditions, the system evolves toward stationary patterns, whereas imposing Dirichlet constraints on the temperature field leads to nonstationary dynamics characterized by persistent temporal evolution of the spatial structures, as illustrated in Movie~S1 in the SI (spot-splitting via pattern-forming instability at $T_{\mathrm{SS}} = 25$). Additionally, we provide numerically simulated spatial patterns for both boundary conditions in the case $B_u \neq B_v$, corresponding to the representative points $B_1$–$B_4$ in Fig.~S1, as shown in Fig.~S2 of the SI.

These results demonstrate that thermal coupling not only tunes instability strength and wavelength but also introduces sensitivity to boundary-imposed thermal constraints, thereby modifying global pattern organization. Notably, this effect has no analogue in purely isothermal reaction-diffusion systems.

\subsection{MODEL--II: Schnakenberg Reaction--Diffusion System}
\subsubsection*{Baseline isothermal instability structure}

To assess the generality of thermally modulated pattern formation, we next consider the classical Schnakenberg reaction-diffusion model, a prototypical activator-inhibitor system that exhibits diffusion-driven (Turing) instability. In dimensionless form, the governing equations are
\begin{align}
\frac{\partial u}{\partial t} &= a - u + u^2 v + \nabla^2 u, \\
\frac{\partial v}{\partial t} &= b - u^2 v + D \nabla^2 v,
\end{align}
where $u$ and $v$ denote the activator and inhibitor concentrations, respectively. The parameters $a$ and $b$ control the feed rates of the two species, and $D$ represents the ratio of diffusion coefficients. The homogeneous steady state is given by
\begin{equation}
u_{\mathrm{ss}} = a + b, 
\qquad 
v_{\mathrm{ss}} = \frac{b}{(a+b)^2}.
\end{equation}
Linear stability analysis about $(u_{\mathrm{ss}}, v_{\mathrm{ss}})$ (see SI) yields the conditions for diffusion-driven instability.
The resulting bifurcation diagram in the $(b,D)$ parameter space is shown in Fig.~\ref{Fig:Schnak_bif}(a), delineating regions of homogeneous stability and Turing instability. Two representative points, $P$ and $Q$, are selected within the homogeneous region but located at different positions relative to the instability boundaries. The corresponding numerical simulations, shown in Figs.~\ref{Fig:Schnak_bif}(b) and (c), confirm that both points yield spatially uniform steady states in the isothermal limit.
This choice allows us to examine how thermal renormalization shifts the instability boundaries and induces transitions from homogeneous states to pattern-forming regimes. In particular, their distinct locations in parameter space enable us to probe how temperature-driven kinetic renormalization can selectively drive the system across instability thresholds.
These results establish the baseline instability structure of the Schnakenberg system and provide a reference for analyzing temperature-induced transitions in the following sections.

\subsubsection*{Thermal renormalization in the Schnakenberg system}

Having established the baseline isothermal stability structure, we now examine how thermal coupling modifies the instability landscape of the Schnakenberg system. As in the CDIMA case, temperature enters through Arrhenius-type prefactors that renormalize the effective reaction kinetics. The temperature-coupled equations are given by
\begin{align}
\frac{\partial u}{\partial t} &= \nabla^2 u + h_u(T)\left(a - u + u^2 v\right), \\
\frac{\partial v}{\partial t} &= D \nabla^2 v + h_v(T)\left(b - u^2 v\right), \\
\frac{\partial T}{\partial t} &= D_T \nabla^2 T 
- h_u(T)\left(a - u + u^2 v\right)
- h_v(T)\left(b - u^2 v\right),
\end{align}
where $D_T$ is the thermal diffusivity and 
$h_u(T) = A_u e^{-B_u/T}$ and $ h_v(T) = A_v e^{-B_v/T} $ are the temperature-dependent kinetic prefactors.

Linear stability analysis about the homogeneous steady state yields modified Hopf and Turing instability conditions in the presence of thermal coupling (see SI for derivation). The Hopf bifurcation condition is given by
\begin{equation}
A_u < - A_v \, e^{\frac{B_u - B_v}{T_{\mathrm{ss}}}}
\left[\frac{-(a+b)^3}{(b-a)}\right],
\end{equation}
while the Turing instability condition is
{
\begin{widetext}
\begin{equation}
\Big[
D A_u e^{-B_u/T_{\mathrm{ss}}} (2u_{\mathrm{ss}}v_{\mathrm{ss}}-1)
- A_v e^{-B_v/T_{\mathrm{ss}}} u_{\mathrm{ss}}^2
\Big]
- 2 \sqrt{
D u_{\mathrm{ss}}^2 A_u A_v
e^{-\frac{(B_u+B_v)}{T_{\mathrm{ss}}}}
}
> 0.
\end{equation}
\end{widetext}}
\begin{figure*}[t]
\centering
\includegraphics[width=0.9\textwidth]{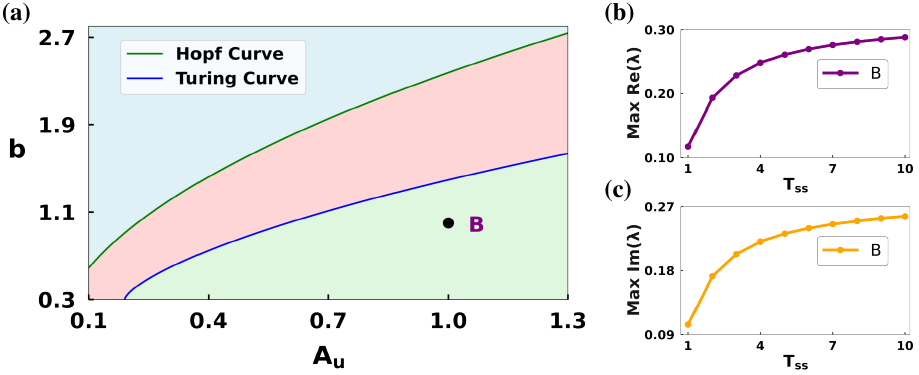}
\caption{
$(a)$ Two-parameter bifurcation diagram of the temperature-coupled Schnakenberg model in the $A_u$--$b$ parameter space at fixed $a = 0.1$, $A_v = 0.15$, and $B_u = B_v = 1$ for a representative homogeneous steady-state point $Q(D=2.5,\; b=1.0)$. Light green denotes the spatiotemporally unstable region, light blue the spatiotemporally stable region, and light pink the spatially stable but temporally unstable region. 
$(b)$ Maximum growth rate $\max \mathrm{Re}(\lambda)$ and $(c)$ $\max \mathrm{Im}(\lambda)$ as a function of temperature evaluated at point $B(1,1)$.
}
\label{Fig:Schnak_bif_Q}
\end{figure*}
To probe the effect of thermal renormalization, we construct bifurcation diagrams in the $(A_u,b)$ parameter space for representative homogeneous steady-state points. Although the system is initially in a homogeneous state in the isothermal limit, thermal renormalization shifts the instability boundaries and can drive the system into pattern-forming regimes.
Further insight is obtained from the dispersion relation. Figures~\ref{Fig:Schnak_bif_P}(b,c) show the variation of the maximum growth rate $\max\{\mathrm{Re}(\lambda)\}$ and the corresponding dominant wavelength as functions of temperature for two representative parameter sets denoted by points $A$ and $A^{\prime}$.
In both cases, $\max\{\mathrm{Re}(\lambda)\}$ increases with temperature, reflecting enhanced reaction kinetics due to Arrhenius scaling. However, the growth rate remains consistently higher for point $A^{\prime}$, which lies closer to the instability boundary. This indicates that proximity to the bifurcation threshold enhances the sensitivity of the system to thermal renormalization.
A similar trend is observed for the dominant wavelength. As temperature increases, the wavelength decreases for both parameter sets, indicating a shift toward finer spatial structures. However, the wavelength remains larger for point $A$ compared to $A^{\prime}$, reflecting differences in the effective distance from the instability boundary.

An analogous behavior is observed for the representative point $B$ in Fig.~\ref{Fig:Schnak_bif_Q}(a). The maximum growth rate $\max\{\mathrm{Re}(\lambda)\}$ increases monotonically with temperature, as shown in Fig.~\ref{Fig:Schnak_bif_Q}(b), indicating progressive destabilization of the homogeneous steady state. However, the resulting dynamical response differs from that observed at points $A$ and $A^{\prime}$, further emphasizing that thermal effects act in conjunction with the underlying bifurcation structure in determining the system’s response.

Overall, these results demonstrate that thermal renormalization systematically enhances diffusion-driven instability and controls wavelength selection. However, in contrast to the CDIMA system, the overall structure of the instability diagram remains qualitatively unchanged. Thermal coupling primarily shifts the location of instability boundaries without introducing additional dynamical regimes, indicating that in the Schnakenberg model temperature acts predominantly as a parametric deformation of the underlying kinetic phase space.

Taken together, these observations show that the effect of temperature depends sensitively on the system’s position relative to the instability boundaries. We now turn to numerical simulations to examine how these thermally induced changes manifest in the nonlinear regime.

\begin{figure*}[t]
\centering
\includegraphics[width=0.995\textwidth]{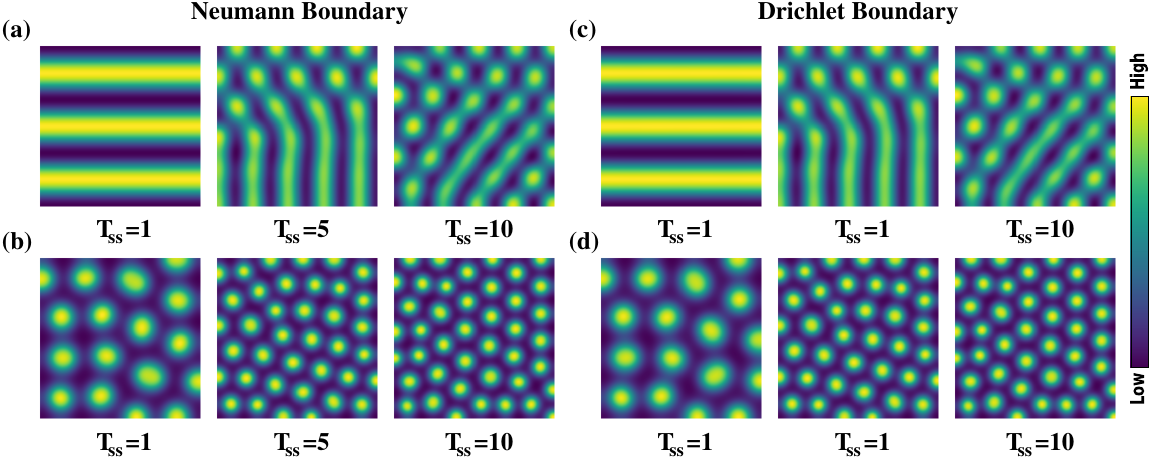}
\caption{
Numerically simulated spatial patterns in the temperature-coupled Schnakenberg model. 
$(a,b)$ Patterns obtained with Neumann boundary conditions, and $(c,d)$ patterns obtained with Dirichlet boundary conditions. 
Panels $(a,c)$ correspond to point $A$, while panels $(b,d)$ correspond to point $A^{\prime}$. Simulations are performed with diffusion coefficients $D_T = 0.01$ and $D = 12.5$ on a square domain of size $1 \times 1$ with spatial discretization $\Delta x = 0.01$ and a time step of $dt=0.002$.
}
\label{Fig:Schnak_patterns}
\end{figure*}

\subsubsection*{Numerical patterns and boundary-condition effects}

Direct numerical simulations provide a visual and dynamical confirmation of the analytical predictions for the temperature-coupled Schnakenberg system. Stationary Turing patterns emerge over a broad temperature range, with representative steady-state configurations under Neumann (zero-flux) boundary conditions for the activator $u$ shown in Fig.~\ref{Fig:Schnak_patterns}(a) and (b) for the parameter sets $A$ and $A^{\prime}$, respectively. In both cases, the patterns evolve from small-amplitude fluctuations and saturate into time-independent spatial structures, consistent with the dispersion-relation analysis, which predicts purely real positive growth rates in the Turing regime. The absence of oscillatory components in the eigenvalue spectrum is reflected in the stationary nature of the resulting patterns. A clear temperature dependence is evident from the spatial snapshots. As temperature increases ($T$ ranges from 1 to 10), the characteristic wavelength decreases, leading to finer spatial features within the same domain. This observation is consistent with the shift of the most unstable mode toward higher wavenumbers at elevated temperatures.

Importantly, qualitative differences between the two parameter sets are also observed. For point $A$, located closer to the Turing bifurcation boundary, the patterns exhibit stripe-like or labyrinthine morphologies ($T=1$), indicative of near-critical mode selection. In contrast, for point $A^{\prime}$, which lies farther from the Turing threshold and closer to the Hopf boundary, the structures are more localized and spot-like structures. This distinction highlights how proximity to different bifurcation thresholds influences not only instability strength but also the resulting pattern morphology under thermal modulation.

\begin{figure}[t]
\centering
\includegraphics[width=0.5\textwidth]{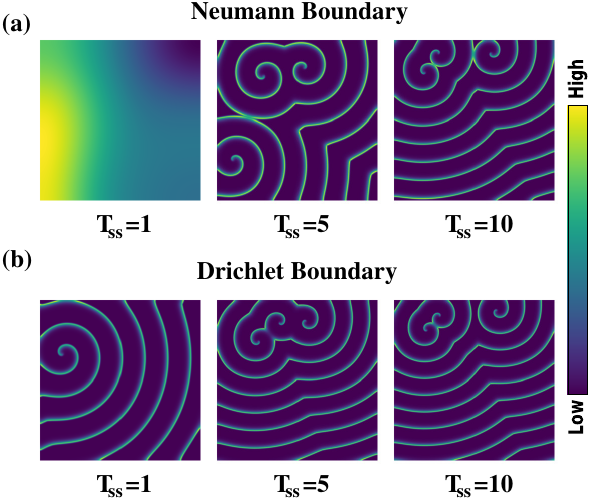}
\caption{
Numerically simulated spatial patterns in the temperature-coupled Schnakenberg model at point $B$: 
$(a)$ Neumann boundary condition and $(b)$ Dirichlet boundary condition. Simulations are performed with diffusion coefficients $D_T = 0.01$ and $D = 2.5$ on a square domain of size $5 \times 5$ with spatial discretization $\Delta x = 0.01$ and a time step of $dt=0.002$.
}
\label{Fig:Schnak_patterns_spairals}
\end{figure}

We next examine the role of temperature in modulating spatiotemporal dynamics at the representative parameter point $B$. Fig.~\ref{Fig:Schnak_patterns_spairals} illustrates the evolution of patterns under different thermal boundary conditions. Under Neumann (zero-flux) boundary conditions [Fig.~\ref{Fig:Schnak_patterns_spairals}(a)], the system remains nearly homogeneous at low temperaturesand develops spatial structure only beyond a threshold. At higher temperatures, coherent structures such as spiral patterns emerge (Movie-S2 in SI), reflecting the amplification of instabilities through thermal renormalization. In contrast, under Dirichlet boundary conditions [Fig.~\ref{Fig:Schnak_patterns_spairals}(b)], spatial structures appear even at lower temperatures, reflecting an earlier onset of instability due to boundary-imposed thermal constraints. As temperature increases further, the patterns become more organized and exhibit stronger boundary influence, leading to more confined and structured morphologies compared to the Neumann case.


\subsection*{Comparison between CDIMA and Schnakenberg systems}
We now compare the effects of thermal coupling in the CDIMA and Schnakenberg reaction–diffusion systems to identify the general features and model-dependent responses arising from temperature-induced kinetic renormalization.

In both systems, temperature enters through Arrhenius-type prefactors that renormalize the effective reaction kinetics. As a result, thermal coupling systematically enhances instability growth rates and shifts the most unstable mode toward higher wavenumbers, leading to a reduction in the characteristic wavelength of the emergent patterns. This behavior is consistently observed in both models, indicating that thermal renormalization provides a general mechanism for controlling instability strength and spatial scale in reaction–diffusion systems.

Despite these common features, the two systems exhibit markedly different responses to thermal coupling at the level of global dynamics and boundary sensitivity. In the CDIMA system, thermal effects lead not only to quantitative shifts in instability thresholds but also to qualitative changes in the dynamical behavior. In particular, the imposition of Dirichlet boundary conditions on the temperature field induces nonstationary dynamics, with persistent temporal evolution of spatial structures. This indicates a strong coupling between thermal gradients and the underlying reaction kinetics, leading to boundary-sensitive pattern selection and dynamical transitions.

In contrast, the Schnakenberg system exhibits a more robust response to thermal modulation. Although thermal renormalization shifts the instability boundaries and modifies growth rates and wavelength selection, the overall structure of the phase diagram remains qualitatively unchanged. The system continues to support stationary diffusion-driven patterns under both Neumann and Dirichlet thermal boundary conditions, and no transition to nonstationary dynamics is observed. This indicates that, in the Schnakenberg model, temperature acts primarily as a parametric deformation of the kinetic phase space rather than inducing new dynamical regimes.
This difference arises from the sensitivity of each system to its position relative to the instability boundaries. In both models, states closer to the Turing threshold exhibit stronger thermal sensitivity, as reflected in enhanced growth rates and distinct pattern morphologies. However, the CDIMA system shows a stronger nonlinear response to such shifts, allowing thermal coupling to qualitatively alter the dynamical state. In contrast, the Schnakenberg system responds more smoothly, with thermal effects modulating pattern characteristics without destabilizing the stationary regime.

Overall, these results demonstrate that while thermal renormalization provides a universal mechanism for tuning instability strength and spatial structure, its impact on global dynamics is strongly model dependent. The interplay between thermal feedback, reaction nonlinearities, and boundary conditions determines whether temperature acts as a simple control parameter or as a driver of qualitative dynamical transitions.

\section{Concluding remarks}

We have investigated the impact of thermal coupling on pattern formation in reaction–diffusion systems by introducing a dynamically evolving temperature field that modulates reaction kinetics through Arrhenius-type prefactors. This framework leads to an effective temperature-dependent renormalization of the underlying reaction rates, providing a systematic mechanism for controlling both instability thresholds and spatial pattern selection.

Across both the CDIMA and Schnakenberg models, thermal coupling is found to enhance instability growth rates and shift the dominant unstable mode toward higher wavenumbers, resulting in finer spatial structures at elevated temperatures. This demonstrates that temperature acts as a robust control parameter for tuning the intrinsic length scale of diffusion-driven patterns.

Despite these common features, the two systems exhibit fundamentally different responses at the level of global dynamics. In the CDIMA model, thermal coupling induces strong sensitivity to boundary conditions and can lead to qualitative changes in dynamical behavior, including the emergence of nonstationary patterns under Dirichlet thermal constraints. In contrast, the Schnakenberg system exhibits a more robust response, with thermal effects primarily shifting instability boundaries and modulating pattern characteristics without altering the stationary nature of the patterns.

These results demonstrate that thermal renormalization provides a general mechanism for tuning instability strength and spatial structure, while its impact on global dynamics remains strongly model dependent. In particular, the interplay between thermal feedback, reaction nonlinearities, and boundary conditions determines whether temperature acts as a simple control parameter or induces qualitative dynamical transitions. More broadly, this work highlights temperature as an active and tunable field in pattern-forming systems.

The present framework can be extended to more complex reaction networks, spatially heterogeneous environments, and externally driven thermal fields, opening new avenues for controlled pattern engineering in chemically reactive and thermally active systems. These findings suggest that thermal control strategies must be tailored to the underlying reaction kinetics, as identical thermal perturbations can lead to fundamentally different dynamical outcomes.

\section*{Supporting Information}
The supporting information provides the detail description of the linear stability analysis mentioned in the main text and includes the supporting figures corresponding to the results discussed in the main text. The supporting video files are also included here.

\section*{Conflicts of interest}
There are no conflicts to declare.

\section*{Acknowledgements}
We greatly acknowledge Indian Institute of Science Education and Research Thiruvananthapuram (IISERTVM), India for providing the support of computing resources. SD acknowledges the institute fellowship from IISERTVM.

\bibliography{reference}

\end{document}